\documentclass[runningheads]{llncs}

\usepackage[T1]{fontenc}
\usepackage[utf8]{inputenc}
\usepackage{amsmath,amssymb}
\usepackage{graphicx}
\usepackage{array}
\usepackage{float}
\usepackage{url}
\usepackage{tikz}
\usetikzlibrary{arrows.meta,positioning,fit,backgrounds,calc}
\usepackage[table]{xcolor}

\definecolor{mffblue}{HTML}{DCEAF7}
\definecolor{mffblueedge}{HTML}{2F5597}
\definecolor{mffgreen}{HTML}{E2F0D9}
\definecolor{mffgreenedge}{HTML}{548235}
\definecolor{mffgray}{HTML}{F2F2F2}
\definecolor{mffgrayedge}{HTML}{666666}
\definecolor{mfforange}{HTML}{FCE4D6}
\definecolor{mfforangeedge}{HTML}{C55A11}

\begin{document}

\title{Math Foundation Formalization Agent:\\
A General-Purpose Approach to Formalizing Mathematical Foundations}

\titlerunning{MFF-Agent}

\author{Jiaxin Xie \and
Guowei Dou \and
Xiang Ji \and
Wensheng Yu}

\authorrunning{J. Xie et al.}

\institute{School of Electronic Engineering,
Beijing University of Posts and Telecommunications, Beijing, China\\
\email{xiejx@bupt.edu.cn, dgw@bupt.edu.cn, jxxxx2001@bupt.edu.cn, wsyu@bupt.edu.cn}}

\maketitle

\begin{abstract}
Mathematical formalization plays an important role in establishing rigorous
and machine-verifiable foundations for mathematical reasoning. With the
development of large language models (LLMs) and interactive theorem provers,
intelligent agents offer a promising approach to automating the formalization
of mathematical theories. This paper presents the Math Foundation
Formalization Agent (MFF-Agent), a general-purpose intelligent agent designed
to support formalization tasks across multiple mathematical domains. The agent
integrates the reasoning capabilities of large language models with the formal
verification capabilities of the Rocq proof assistant. It incorporates task
classification, tool-assisted execution, verification feedback, and state-based
task management to support the generation, refinement, and verification of
formal proofs. Unlike an approach focused on a single mathematical theory, the
agent is examined through case studies involving Morse--Kelley axiomatic set
theory (MK), real numbers, propositional logic, and equivalences of the axiom of
choice. These cases investigate its applicability across mathematical
foundations and distinguish source-level completion, interactive verification,
and reproducible compilation. The results demonstrate the feasibility of a
unified operational framework for heterogeneous foundation-level tasks and
clarify the gap between such a framework and a fully integrated mathematical
foundation.

\keywords{Formalized mathematics \and Autonomous agents \and Mathematical foundations \and Autoformalization.}
\end{abstract}

\section{Introduction}

Formalized mathematics represents mathematical definitions, statements, and
proofs in formal languages that can be mechanically verified by proof
assistants, providing a rigorous foundation for reliable mathematical reasoning
and computer-assisted mathematics~\cite{Koutsoukou2021}. Over the past decades,
interactive theorem provers (i.e., proof assistants) such as Rocq (formerly
Coq)~\cite{Rocq}, Isabelle~\cite{Isabelle}, and Lean~\cite{Lean} have become
important tools in formal verification and computational logic~\cite{Coq2}.
Landmark achievements, including the machine-verified proofs of the four color
theorem~\cite{fourcolor} and the Feit--Thompson theorem~\cite{feit-thompson},
have demonstrated the potential of formal methods for establishing mathematical
results. Recent advances in large language models (LLMs) have created new
opportunities for automating mathematical reasoning and proof development.
Integrating LLMs with proof assistants offers a promising approach to generating
proof scripts, interpreting verification feedback, and refining formal proofs
through iterative interaction~\cite{AIformal,Tao,Buzzard,CoqPilot,Numina}.
However, the formalization of mathematical foundations raises requirements that
are not captured by isolated proof generation. A useful agent must handle
heterogeneous requests, maintain long-lived interactions with a proof
assistant, reuse declarations across related files, and determine whether the
requested task has actually been completed rather than merely whether some
source file compiles.

Recent research has explored diverse agent-based approaches to formal
mathematical reasoning and proof development. CoqPilot integrates large
language models (LLMs) with Coq to generate proof candidates and refine them
using feedback from the proof assistant~\cite{CoqPilot}, while
Numina-Lean-Agent investigates the use of a general-purpose coding agent for a
broad range of formal mathematics tasks~\cite{Numina}. Prover Agent further
coordinates informal reasoning, formal proof construction, and feedback from
Lean, using auxiliary lemma generation to facilitate the discovery of proof
strategies for challenging problems~\cite{ProverAgent}. In a complementary
direction, LeanAgent adopts a lifelong learning framework to improve
generalization across mathematical domains and accumulate knowledge from
successive formalization tasks~\cite{LeanAgent}. More recently,
\emph{Formalizing Mathematics at Scale} presents AutoformBot, a multi-agent
system that coordinates large-scale formalization through formal verification
tools, dependency-aware task scheduling, and collaborative version
control~\cite{AutoformBot}. Together, these systems demonstrate substantial
progress in proof generation, iterative refinement, knowledge reuse, and
large-scale formalization. Their primary emphasis, however, remains proof search
performance, long-horizon proof development, or a particular formalization
pipeline. In contrast, the present work does not aim primarily to increase the
success rate on an isolated theorem benchmark. Its focus is the operational
problem of supporting heterogeneous formalization tasks and cross-file
foundation artifacts within one execution model, while deciding completion from
evidence returned by the proof assistant and task-specific checks.

\emph{Research objective.} We develop the Math Foundation Formalization Agent
(MFF-Agent), a general-purpose agent for formalizing mathematical foundations
in Rocq. MFF-Agent processes different formalization requests, invokes Rocq
through persistent proof-state interaction, adapts its reasoning and context
management to task progress, and determines task status from explicit
verification conditions rather than from the LLM's self-report alone.

\emph{Main contributions.} This paper makes the following contributions:

\begin{enumerate}
\item \textbf{Unified operational framework for foundation-level tasks.}
MFF-Agent supports heterogeneous tasks, including theorem proving, file
creation, modification, analysis, and compilation, across different
foundational topics.

\item \textbf{Adaptive formalization process.} The agent combines dual-mode
reasoning, dynamic interaction with Rocq, and context management to support
iterative proof development and long-running formalization tasks.

\item \textbf{Reliable task completion verification.} A state-based control
mechanism and task-specific completion checks distinguish successful
compilation from the actual completion of the requested formalization task.
\end{enumerate}

The remainder of this paper is organized as follows. Section 2 defines the
success criteria for formalization tasks and introduces the evaluation harness.
Section 3 presents the MFF-Agent framework, including its overall workflow,
dual-mode reasoning mechanism, dynamic Rocq session management,
verification-driven task completion control, and context pruning. Section 4
reports the formalization results across the selected mathematical domains and
discusses the available verification evidence and current limitations. Finally,
Section 5 concludes the paper and outlines directions for future work.

\section{Success Criteria and Evaluation Harness}
\label{sec:criteria}

Before presenting the overall architecture of the Math Foundation
Formalization Agent, we first define the criteria for determining task
completion. During preliminary experiments, several types of false-positive
completion reports were observed. For example, the large language model (LLM)
reported success even though some target theorems remained unproved; a test
file generated during the task compiled successfully, leading the agent to
incorrectly conclude that the original task was complete; or only part of a
multi-theorem task had been completed. These cases indicate that neither the
LLM's self-reported status nor the successful compilation of an arbitrary file
is sufficient to establish task completion.

To address these issues, we introduce the following design measures:
\begin{enumerate}
\item \emph{Task-specific completion criteria.} The agent distinguishes among
five task types: proof, creation, modification, analysis, and compilation.
Each type is evaluated according to its corresponding completion requirements.

\item \emph{Target-file verification.} For proof tasks, compilation success
must be associated with the intended target file, rather than a separately
generated test file. The agent also checks whether the requested target
theorems have been completed.

\item \emph{Proof-completeness checks.} Constructs such as
\texttt{Admitted}, \texttt{admit}, and \texttt{Abort} are treated as
indicators of unfinished proofs. These are distinguished from explicit axioms
that belong to the underlying formal theory.

\item \emph{Programmatic success determination.} The agent reports a task as
complete only when the applicable checks are satisfied, rather than relying
solely on the LLM's self-reported result.
\end{enumerate}

Together, these measures provide explicit operational criteria for task
completion and reduce false-positive success reports across different
mathematical formalization tasks.

\section{The MFF-Agent Framework}

\begin{figure}[H]
    \centering
    \includegraphics[width=0.66\textwidth]{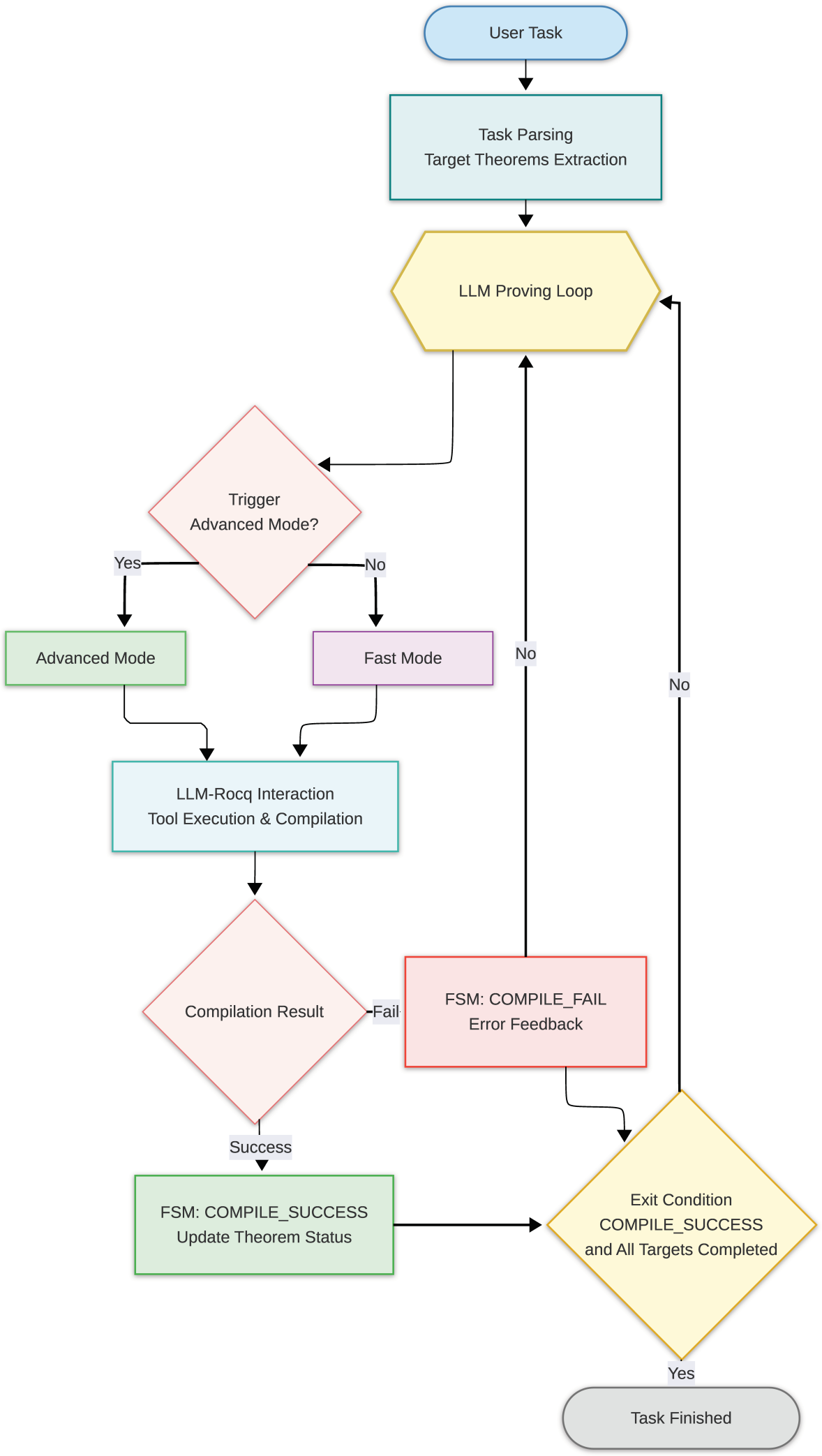}
    \caption{The framework of the agent}
    \label{fig1}
\end{figure}

The overall architecture of the proposed MFF-Agent is illustrated
in Fig.~\ref{fig1}. The agent integrates the reasoning capability of LLMs with
the verification capability of the Rocq proof environment. Although theorem
proving is its primary use case, the implementation is not restricted to proof
generation: it distinguishes among proof, file-creation, modification,
analysis, and compilation tasks. The common workflow consists of task parsing,
iterative reasoning, tool execution, and result verification.

The agent first receives the user request and constructs a structured task
specification containing the task type, source and target files, target
theorems when applicable, a natural-language description, and completion
criteria. It then enters the main execution loop, selects an appropriate
reasoning mode, invokes the LLM, executes file or Rocq tools, and uses the
returned compiler or proof-state information in the next iteration. The
completion path depends on the task type: proof tasks require compilation and
target-proof checks, whereas creation, modification, analysis, and compilation
tasks use their respective completion policies.

The loop terminates when a task-specific check succeeds, when an early-stop or
clarification condition is reached, or when the iteration budget is exhausted.
A task-summary file is generated for every one of these outcomes, including
failed and incomplete executions, before the agent returns to the
waiting-for-user state.

%%%%%%%%%%3.1%%%%%%%%%%
\subsection{Dual-Mode Switching Mechanism}

During automated formalization, different tasks require different levels of
reasoning capability and contextual information. Excessive context increases
token consumption and reasoning overhead for routine operations, whereas
insufficient context may prevent the model from finding an effective proof
strategy. To balance these costs, the agent employs Fast Mode and Advanced
Mode.

The agent starts in Fast Mode. Its prompts contain the current task
specification, completion requirements, the latest compiler feedback, and a
task-specific action hint, without including the accumulated failure history or
a planner-generated proof plan. The model can directly create or edit Rocq
source files and invoke compilation to refine the result. Fast Mode is a prompt
and context configuration rather than a restriction on the available tool set;
the same Rocq tools remain callable in both modes.

The agent enters Advanced Mode under any of three conditions: ten consecutive
compilation failures; a newly reached multiple of five completed target
theorems; or an estimated context size above 500,000 tokens when no previous
compression exists or the current estimate exceeds the previous post-compression
size by more than 30\%. Advanced Mode exposes the latest proof state, recent
failed attempts, compiler-error categories, and strategy hints. The Planner is
invoked only for the ten-failure trigger. The theorem-progress and
context-length triggers enter Advanced Mode for context consolidation but do
not generate a new Planner result.

\begin{figure}[H]
    \centering
    \includegraphics[width=0.68\textwidth]{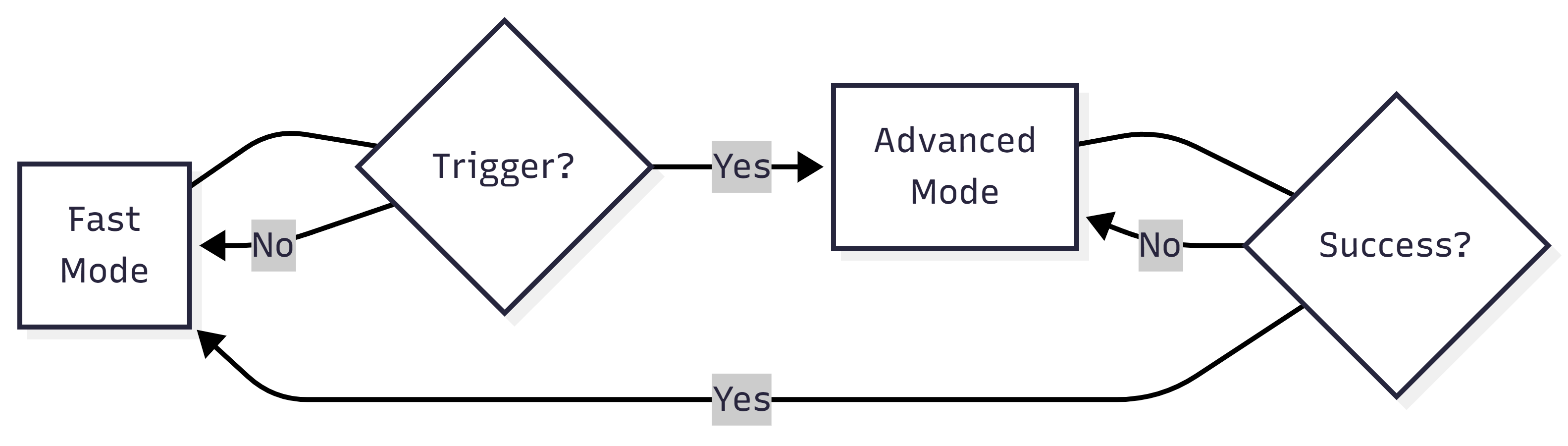}
    \caption{Workflow of the dual-mode switching mechanism}
    \label{fig2}
\end{figure}

As shown in Fig.~\ref{fig2}, the mode changes dynamically during execution.
When a compilation attempt succeeds, the agent immediately returns to Fast
Mode. This transition is based on compilation success, not on full task
acceptance: if an additional task-specific check still fails, the agent remains
in the execution loop and applies the corresponding incomplete-task feedback.

%%%%%%%%%%3.2%%%%%%%%%%
\subsection{Dynamic Rocq Session Management}

The interaction between the agent and the Rocq environment is implemented by a
\texttt{CoqSession} object. A straightforward implementation would launch a
new \texttt{coqtop} process and reload the entire proof file for each request.
The proposed session instead retains a \texttt{coqtop} subprocess while the
source remains unchanged.

Any successful file-creation or replacement operation invalidates the existing
session because its loaded state no longer agrees with the source file. The
next proof-state request starts a fresh session for the current file; requests
for the same unchanged file reuse the existing session. This invalidation rule
prevents stale goals from being associated with newly edited source code.

\begin{figure}[htbp]
    \centering
    \includegraphics[width=0.54\textwidth]{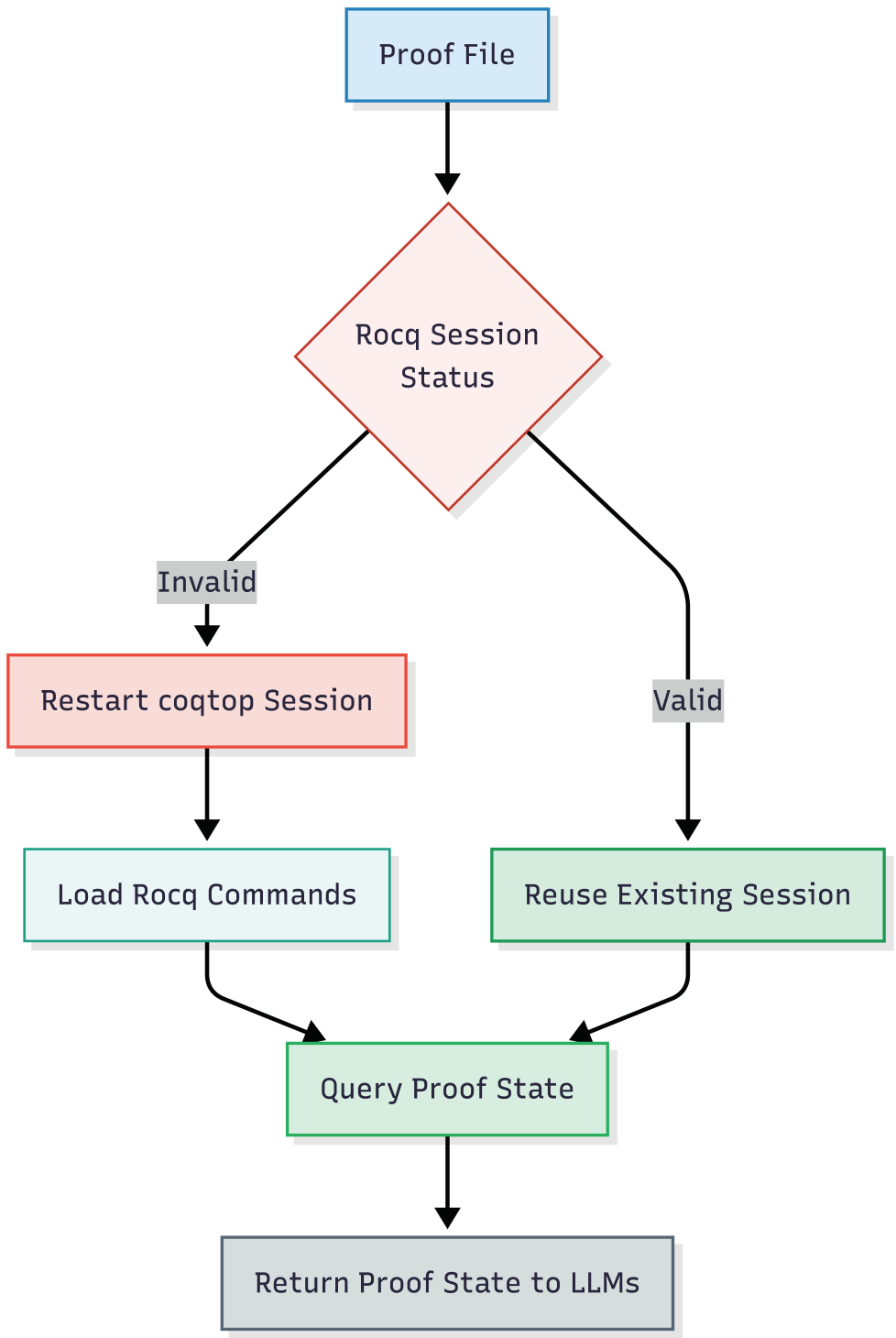}
    \caption{Workflow of dynamic Rocq Session management}
    \label{fig3}
\end{figure}

The proof-state retrieval path first uses a cached or newly executed
\texttt{coqc} result to determine the failing line. If a reliable line number
is available, the agent starts a session and sends only complete Vernac commands
from the valid prefix before that line. Intermediate outputs are not read for
each command; after the prefix is loaded, the agent issues \texttt{Show.} once
to obtain the active goal and hypotheses. If no reliable error line can be
determined, the agent does not load the whole file into \texttt{coqtop}; it
returns a compiler diagnostic, or reports that no active proof can be queried
when compilation has already succeeded. The resulting proof-state and compiler
information is then returned to the LLM for the next reasoning step.

%%%%%%%%%%3.3%%%%%%%%%%
\subsection{Verification-driven Task Completion Control}
\label{sec:runtime-completion}

Section~\ref{sec:criteria} defines what acceptance means and how it is
evaluated. This subsection describes the runtime mechanism that decides whether
the current execution may stop successfully. Completion control combines a
finite-state machine with a task-specific completion manager.

The Agent FSM defines six states: INIT, EXECUTING,
COMPILE$\_$SUCCESS, COMPILE$\_$FAIL, TASK$\_$SUCCESS, and
TASK$\_$FAIL. A task starts in INIT, and any source edit moves the machine to
EXECUTING while invalidating the previous compiler result. A compiler attempt
then moves the machine to COMPILE$\_$SUCCESS or COMPILE$\_$FAIL.
TASK$\_$SUCCESS is entered after a successful task-level check.
TASK$\_$FAIL is currently a reserved final state: the implementation records
failed, incomplete, and input-required outcomes in a separate
\texttt{final\_status}, but does not transition the FSM to TASK$\_$FAIL.

The \texttt{TaskCompletionManager} dispatches according to the task type. For a
proof task, \texttt{TaskCompletionChecker} locates the target file and theorem
range, requires compilation, and checks each requested declaration for a proof
block containing \texttt{Proof.} and ending with \texttt{Qed.} or
\texttt{Defined.}. It reports missing proof structures and \texttt{Admitted.};
the current proof checker does not reject an explicitly introduced axiom in a
non-target declaration. For creation tasks, the manager requires an existing,
nonempty, compiling file, checks general unfinished markers and any named
target proofs, and then invokes an LLM goal check. Modification tasks require
compilation, a changed file when a baseline was captured, the same unfinished
marker check, and an LLM goal check; they do not perform a separate target-proof
completeness scan. Compilation tasks require only a successful compiler result
and do not inspect unfinished proof markers. Analysis tasks skip compilation
and require a substantive textual diagnosis covering a cause, a location, and
a repair suggestion.

The completion policies are not yet completely uniform. The general
\texttt{\_find\_admitted} helper used for creation and modification rejects
\texttt{Axiom} and \texttt{Conjecture} together with unfinished proof markers,
whereas the proof-task checker searches only for \texttt{Admitted.} inside the
target proof. This difference must be considered when interpreting runtime
acceptance.

When the LLM stops calling tools, analysis tasks are checked directly; other
task types first require COMPILE$\_$SUCCESS and then invoke the corresponding
completion check. A failed check produces task-specific feedback and the loop
continues. Three consecutive no-tool responses without successful completion
terminate the current run with a failed status. The runtime path checks
compilation and locally implemented task conditions; transitive axiom-footprint
inspection and semantic-fidelity review remain part of the evaluation and audit
layer in Section~\ref{sec:criteria}. Thus, runtime success should not be
interpreted as passing those additional gates.

%%%%%%%%%%3.4%%%%%%%%%%
\subsection{Dynamic Context Pruning}

During long-running tasks, the interaction history continuously accumulates.
The agent therefore compresses messages when Advanced Mode is active and the
conversation exceeds 25 messages. Scale- and context-triggered Advanced Mode
entries invoke the compression immediately; failure-triggered entries use the
regular Advanced Mode compression. Normal compression retains at most 20
messages and six recent messages. The 30\% post-compression growth condition
controls re-entry into Advanced Mode after a token-threshold trigger; it is not
a universal condition for every compression operation.

The compressor preserves the system message and the message in the second
position as the initial user message, then adds a compact summary and recent
messages. The summary records the working file, counts of compilation and
replacement attempts, the latest compilation error, and the latest observed
proof state. Long errors, proof states, and the initial user message are
truncated. Because the conversation history persists across tasks, the second
message can belong to an earlier task; consequently, later compressions may
retain an older task description rather than the newest one.

Recent-message deduplication removes repeated compilation failures and repeated
proof-state results, retaining the latest occurrence of each. It does not
deduplicate successful compilation results. The failure history is separately
limited to three attempts. These attempts are injected into the expanded
context when the ten-failure threshold triggers Advanced Mode, but they are not
stored as a guaranteed field of the compressed-summary object. If older
Advanced Mode messages fall outside the retained recent window, their detailed
edit history may therefore be lost, although the latest compiler error and
proof state remain in the summary.

This combination of compression, selective deduplication, and bounded failure
history keeps the prompt manageable during long executions. The implementation
does not yet guarantee that every historical failure or the most recent task
description survives across task boundaries, which is an important limitation
when interpreting long-horizon behavior.

\section{Results and Analysis}
\label{sec:results}

This section examines MFF-Agent as an intelligent agent for the formalization
of mathematical foundations rather than as a collection of isolated Rocq
libraries. We ask what foundational work has been completed, whether the
results form reusable dependency chains, what assumptions support their
acceptance, and where the current artifact stops short of a unified foundation.
The current source snapshot was checked locally with Coq 8.13.2; the earlier
agent experiments used Rocq 9.0.1. Because the versions differ, source-level
completion, interactive loading, and reproducible compilation are reported as
separate forms of evidence.

\subsection{Completed Foundation Formalization Work}

The current artifact contains nine core Rocq source files comprising 24,379
lines and 647 theorem-like declarations, excluding the duplicate copy of
\texttt{MK.v}. The directory-local \texttt{RAxiom/MK.v} is treated as a build
dependency rather than an additional foundation source. Table~\ref{tab:artifact-results}
summarizes the formalization scope and representative results across the four
domains.

\begin{table}[H]
\caption{Formalization scope and representative results across four
mathematical-foundation domains.}
\label{tab:artifact-results}
\centering
\scriptsize
\setlength{\tabcolsep}{3pt}
\renewcommand{\arraystretch}{1.14}
\begin{tabular}{@{}p{2.1cm}p{5.0cm}p{4.6cm}@{}}
\rowcolor{mffgray}
\textbf{Domain} & \textbf{Formalized content} & \textbf{Representative results} \\
\hline
MK set theory &
Classes and sets; relations and functions; ordinals, cardinals, and finite sets &
The \texttt{MKT4}--\texttt{MKT181b} family, including algebraic, order, and
cardinality results \\
\hline
Choice and equivalents &
Finite-characteristic classes, maximal and minimal members, choice functions,
partial and total orders, chains, and well-orders &
Tukey's lemma, Hausdorff's maximal principle, the maximal principle, Zorn's
lemma, \texttt{WellOrder\_AC}, and \texttt{Zermelo\_AC} \\
\hline
Real numbers &
Ordered field and completeness; supremum and infimum; $\mathbb{N}$,
$\mathbb{Z}$, $\mathbb{Q}$; irrationals; absolute value and distance &
\texttt{Sup\_Corollary}, \texttt{Inf\_Corollary},
\texttt{Existence\_of\_irRational\_Number}, and
\texttt{Real\_struct\_isomorphic} \\
\hline
Propositional logic &
Formula syntax, derivability, truth semantics, proof systems, and
meta-theoretic consistency &
\texttt{soundness\_PC}, \texttt{consistence\_PC}, and \texttt{complete} \\
\end{tabular}
\end{table}

The MK layer provides the broadest foundation. It defines classifiers,
membership, set and class operations, ordered pairs, Cartesian products,
relations, functions, well-orders, ordinals, the set of natural numbers,
cardinal numbers, and finite sets, and then develops the numbered \texttt{MKT}
results. For MFF-Agent, this layer supplies a stable vocabulary that later tasks
can reuse instead of regenerating definitions locally. Its 206 proof-bearing
declarations therefore represent both formalized mathematics and a reusable
working context for subsequent agent tasks. The MK assumptions are collected in
a record containing \texttt{AI}--\texttt{AIX}, while classical logic is supplied
separately through \texttt{classic}.

The choice layer is the clearest example of multi-step foundation work. It
reuses the MK vocabulary to formalize finite-characteristic classes, maximal
and minimal members, choice functions, partial and total orders, chains,
well-orders and initial segments. It then follows the dependency chain
\texttt{Basic\_Definitions} $\rightarrow$ \texttt{Tukey} $\rightarrow$
\texttt{Maximal\_Principle} $\rightarrow$ \texttt{Wellordering} and
\texttt{Zermelo}, proving Tukey's lemma, Hausdorff's maximal principle, Zorn's
lemma, the well-ordering result, and Zermelo's theorem. The corresponding
passages back to a choice function are provided by \texttt{WellOrder\_AC} and
\texttt{Zermelo\_AC}. This case demonstrates the agent's intended ability to
work across linked files and proof obligations whose meaning depends on earlier
foundational declarations.

The real-number work is organized as an axiomatic theory followed by a
structural comparison. From an ordered field with completeness, the artifact
derives arithmetic and order consequences, equivalent formulations of
supremum and infimum, constructions of $\mathbb{N}$, $\mathbb{Z}$ and
$\mathbb{Q}$, rational density, the existence of an irrational number,
Archimedean properties, absolute value and distance. The isomorphism layer
then defines preservation of addition, multiplication and order together with
bijectivity, and proves that any two structures satisfying the same real-axiom
interface are isomorphic. The 339 real-number declarations thus represent both
an axiomatic development and a structural classification result, although their
current build status is treated separately below.

The propositional-logic layer addresses a meta-theoretic foundation and is
currently independent of the MK encoding. It defines formulas, derivability,
truth assignments, tautologies and semantic inference, then proves the basic
proof-system laws, the deduction theorem, soundness, consistency, and
completeness. Auxiliary cardinality results support the completeness argument.
This case shows that the same agent workflow can be applied when the object of
formalization is a proof calculus rather than a mathematical structure built
inside MK.

\subsection{Verification Evidence and Trust Boundary}

The verification evidence is interpreted at three levels. Source-level
completion means that a requested declaration contains a complete proof body
with no unresolved proof marker. Interactive verification means that the source
can be loaded into \texttt{coqtop}, so the proof checker accepts the file in the
current environment. Reproducible compilation means that the required imports
are present and that a current \texttt{.vo} artifact or a clean compiler run can
be obtained. These levels are distinct: a complete source file is not yet a
self-contained build artifact.

The existing MK, \texttt{Basic\_Definitions}, \texttt{Maximal\_Principle} and
\texttt{Tukey} compilation artifacts load successfully. Querying the assumption
footprint of representative theorems in this group returns \texttt{classic},
\texttt{MK\_Axiom}, and the MK primitives \texttt{Class}, \texttt{In} and
\texttt{Classifier}. The files \texttt{Wellordering.v}, \texttt{Zermelo.v} and
\texttt{L\_system\_Logic.v} execute successfully when loaded into
\texttt{coqtop}, although their directories do not contain current \texttt{.vo}
outputs. This establishes that the full choice chain from Tukey's lemma through
the maximal principle, Zorn's lemma, well-ordering and Zermelo's theorem is
accepted by the checker, even though a persistent build workflow is not
preserved in the snapshot.

The real-number foundation is now buildable within its own directory.
\texttt{RAxiom/MK.v} supplies the required MK interface, and both
\texttt{RAxiom/MK.vo} and \texttt{Real\_Axioms1.vo} are current. The complete
isomorphism source also loads successfully into \texttt{coqtop}; its main
theorem, \texttt{Real\_struct\_isomorphic}, has the assumption footprint
\texttt{classic}, \texttt{MK\_Axiom}, and \texttt{MKS}. The 113 declarations in
\texttt{Real\_Axioms1.v} therefore have a reproducible compilation artifact,
while the 226 isomorphism declarations are accepted at source and interactive
levels but do not yet have a retained \texttt{.vo}. This distinction separates
interactive verification from a packaged build artifact.

The trust boundary also determines what the agent has actually established. The
MK record contains choice as \texttt{AIX}; therefore, the choice layer proves
interderivability on an AC-enabled MK base rather than deriving choice from a
base without that principle. The logic soundness proof is closed under the
global Coq context, whereas the completeness proof relies on
\texttt{epsilon\_statement} and \texttt{classic}. At run time, MFF-Agent
organizes these jobs through typed task specifications, a persistent Rocq
session, an execution FSM, and task-specific completion checks. These mechanisms
are exercised by the structure of the artifact, but the snapshot contains no
per-task prompts, tool traces, token counts, or repeated-run logs. We therefore
make no claim about solve rates, speedups, or mechanism ablations from this
chapter.

\subsection{Cross-Foundation Analysis and Limitations}

The artifact shows that MFF-Agent is intended to operate across distinct levels
of mathematical foundation rather than only repair isolated proofs. The
MK-to-choice development exhibits genuine vertical reuse: the later declarations
consume definitions and results from the MK base instead of restating them. The
choice case additionally shows how one formal vocabulary can support several
equivalent organizing principles. The real-number isomorphism shows a different
kind of foundation result: the axiom interface identifies an ordered field up to
isomorphism rather than describing only one accidental model. The logic case
extends the workflow to a meta-theoretic foundation, where the objects being
formalized include formulas, derivations and semantic consequence.

Operational genericity and foundation integration are nevertheless different
claims. MFF-Agent supports heterogeneous tasks across these cases, but the
current dependency structure is not yet a single unified foundation.
Propositional logic imports only Coq's standard libraries and does not build on
the MK source. The real-number development is self-contained within
\texttt{Code/RAxiom} through its local \texttt{MK.v}, while the choice
development reuses another local MK copy. The tree contains two byte-identical
\texttt{MK.v} files and a third, non-identical MK module for the real-number
build. These issues do not prevent the individual developments from being
built, but they prevent the collection from being presented as one canonical
dependency closure.

The agent-level boundary is equally important. The runtime completion
controller enforces compilation and locally implemented target checks, while
transitive axiom-footprint analysis and semantic-fidelity review remain part of
the evaluation process rather than a fully integrated runtime certification.
Completion policies also differ across task types, and the Code snapshot does
not preserve enough execution history to quantify agent performance.

The strongest conclusion supported by the present results is therefore
twofold. First, MFF-Agent demonstrates a unified operational approach to
formalizing several mathematical foundations, with substantial source-level
work and several mechanically accepted proof chains. Second, the artifact has
not yet reached the stronger goal of a fully reproducible, uniformly audited
mathematical foundation. Establishing that goal requires a single canonical
source tree, explicit import and build manifests, complete dependency closure,
transitive axiom checking, semantic review, and task-level execution records.

\section{Conclusion}

This paper presented MFF-Agent, an LLM-driven agent for formalizing
mathematical foundations in Rocq. The framework combines typed task processing,
persistent proof-state interaction, adaptive reasoning and context management,
and task-specific completion control. It has been examined on Morse--Kelley set
theory, the axiom of choice, real-number theory, and propositional logic. These
cases demonstrate operational breadth and concrete dependency reuse, especially
in the MK-to-choice chain and the real-number isomorphism development, while
distinguishing source-level completion from interactive and packaged
verification.

The current results demonstrate the feasibility of a unified execution
framework across mathematical-foundation domains. Building on this foundation,
future work will establish a complete and unified mathematical foundation,
strengthen task completion and trust verification, consolidate the formalized
source and build infrastructure, and extend MFF-Agent to broader application
areas in mathematics and formal reasoning. The code and accompanying artifacts
are available at
\url{https://github.com/jiaxinxie2003/Math-Foundation-Formalization-Agent}.

\subsubsection{Acknowledgements}
This work is supported by the National Natural Science Foundation of China
under Grant 62476028.

\end{document}